\documentclass[preprintnumbers,nofootinbib,twocolumn]{revtex4}

\usepackage{amsmath,bbm,latexsym,url,color}

\newcommand*{\tvec}[1]{\ensuremath{\boldsymbol{\mathrm{#1}}}}           
\newcommand*{\trans}{\mathrm{T}}                     
\DeclareMathOperator{\re}{Re}
\DeclareMathOperator{\tr}{Tr}
\DeclareMathOperator{\im}{Im}

\numberwithin{equation}{section}

\newcommand{\nc}{\newcommand}
\nc{\non}{\nonumber}
\nc{\hc}{\hbox {H.c.}} 
\nc{\noi}{\noindent}
\nc{\barx}{\bar{x}}
\nc{\pbarn}{\;\hbox {pb}}
\nc{\fbarn}{\;\hbox {fb}}
\nc{\lsp}{\;\;\;\;\;}
\nc{\Lsp}{\;\;\;\;\;\;\;\;\;\;}  
\nc{\LLsp}{\lspace \lspace}
\nc{\lra}{\longrightarrow}
\nc{\llra}{\longleftrightarrow}

\nc{\beq}{\begin{equation}}  \nc{\eeq}{\end{equation}}
\nc{\bea}{\begin{eqnarray}}  \nc{\eea}{\end{eqnarray}}
\nc{\baa}{\begin{array}}     \nc{\eaa}{\end{array}}
\nc{\bit}{\begin{itemize}}   \nc{\eit}{\end{itemize}}
\nc{\ben}{\begin{enumerate}} \nc{\een}{\end{enumerate}}
\nc{\bce}{\begin{center}}    \nc{\ece}{\end{center}}
\nc{\bpm}{\begin{pmatrix}}   \nc{\epm}{\end{pmatrix}}
\nc{\bvt}{\begin{verbatim}}  \nc{\evt}{\end{verbatim}}
\def\lsim{\mathrel{\raise.3ex\hbox{$<$\kern-.75em\lower1ex\hbox{$\sim$}}}}
\def\gsim{\mathrel{\raise.3ex\hbox{$>$\kern-.75em\lower1ex\hbox{$\sim$}}}}
\def\lcal{{\cal L}}
\nc{\vp}{\varphi}
\nc{\hp}{\hat\phi}
\nc{\tp}{\tilde\phi}
\nc{\for}{\;\;{\rm for}\;\;}
\nc{\then}{\;\;{\rm then}\;\;}
\nc{\aaa}{\;\;{\rm and}\;\;}

\nc{\p}{\partial}

\def\tr{ \hbox{Tr}}
\def\det{\hbox{det}}
\def\re{{\bf Re}}
\def\im{{\bf Im}}

\def\mati{{\mathbbm1}}

\def\vv{{\bf v}}

\def\su#1{SU(#1)}

\definecolor{purple}{rgb}{.6,0.4,1}
\definecolor{rose}{rgb}{0.93,0.8,0.8}

\begin{document}

\title{Note on Custodial Symmetry in the Two-Higgs-Doublet Model}


\preprint{IFT-10-12}
\preprint{BI-TP-2010/35}
\preprint{UCRHEP-T499}

\author{B. Grzadkowski}
\email[]{Bohdan.Grzadkowski@fuw.edu.pl}
\affiliation{Institute of Theoretical Physics, University of Warsaw, 
Ho\.za 69, PL-00-681 Warsaw, Poland}
\author{M.~Maniatis}
\email[]{Maniatis@physik.uni-bielefeld.de}
\affiliation{Institute for Theoretical Physics, University of Bielefeld,
33615 Bielefeld, Germany}
\author{Jos\'e Wudka}
\email[]{jose.wudka@ucr.edu}
\affiliation{Department of Physics, University of California,
Riverside CA 92521-0413, USA}

\begin{abstract}
We present a simple and transparent method to
study custodial symmetry in the Two-Higgs-Doublet Model. 
The method allows to formulate the basis independent, sufficient and 
necessary, conditions for the custodial symmetry of the scalar potential.
The relation between the custodial transformation and CP
is discussed and clarified.

\end{abstract}


\maketitle

\section{Introduction}

The $\rho$~parameter is experimentally 
measured as~\cite{pdg2010}
\begin{equation}
\label{exrho}
\rho =1.0008^{+0.0017}_{-0.0007}\,,
\end{equation}
where at tree-level 
$\rho \equiv {m_W^2}/{\cos^2 (\theta_W)/m_Z^2}$
with $\theta_W$ the Weinberg angle and
$m_W$ and $m_Z$ the electroweak gauge boson masses.

In the Standard Model~(SM) with a Higgs sector
consisting of one
Higgs doublet $\phi$ there is an extra symmetry of the Higgs potential
\begin{equation}
\label{potSM}
V_{\text{SM}} = -\lambda (\phi^\dagger \phi) + \mu (\phi^\dagger \phi)^2 \;.
\end{equation}
which is responsible for $\rho \approx 1$;
because of its role in insuring
small corrections to $ \rho $ this symmetry of $ V_{\text{SM}}$ is commonly
called a custodial symmetry (CS)~\cite{Sikivie:1980hm} . Its
form can be made manifest by decomposing the complex Higgs-doublet
into the real components $\phi_1$, $\phi_2$, $\phi_3$, $\phi_4$, 
\begin{equation}
\phi= 
\begin{pmatrix}
\phi^+\\ \phi^0
\end{pmatrix}
=
\begin{pmatrix}
\phi_1 + i \phi_2\\ 
\phi_3 + i \phi_4
\end{pmatrix}.
\end{equation}
We then find immediately 
$(\phi^\dagger \phi)= \phi_1^2+\phi_2^2+\phi_3^2+\phi_4^2$,
hence, the potential is invariant 
under $SO(4) \sim SU(2)_L \times SU(2)_R$,
($\sim$ means that both sides have the same Lie algebra),
and the CS is then the diagonal $\su2$ subgroup.

In order to study the CS it is convenient to introduce the following
matrix
\begin{equation}
\label{defMSM}
M \equiv 
\begin{pmatrix} i \sigma^2 \phi^*, & \phi \end{pmatrix} =
\begin{pmatrix} 
\phi^{0\,*} & \phi^+\\
-\phi^-    & \phi^0
\end{pmatrix}\;.
\end{equation}
Then $M$ transforms under $SU(2)_L \times SU(2)_R$ as
\begin{equation}
\label{CStransSM}
M \rightarrow L M R^\dagger\;, 
\end{equation}
where $L,R \in SU(2)_{L,R}$ respectively. 
In the SM one can assume without loss of generality that
the vacuum expectation value of $\phi$ is real, whence, after
spontaneous symmetry breaking (SSB), $\langle M \rangle\,\propto \,\mati_2$, so
$SU(2)_L \times SU(2)_R$ is broken to $SU(2)_{\rm diag}$ (the diagonal subgroup) and 
it is the invariance of $|D_\mu\phi|^2$
with respect to 
$\su2_{\rm diag}$ that insures $\rho=1$ at 
tree-level \cite{Sikivie:1980hm, Peskin:1995ev}.
The term {\em custodial symmetry}
is reserved in the literature for the $SU(2)_{\rm diag}$ transformations
under which both would-be Goldstone bosons and the corresponding
gauge bosons transform as triplets.
In this work we will also refer to
{\em  custodial transformations}~(CT) as those generated by the full group 
 $SU(2)_L \times SU(2)_R$.
We will always assume that the vacuum respects the diagonal $SU(2)$, the CS.

Now we can write
\begin{equation}
\label{scalarM}
(\phi^\dagger \phi)  = \frac{1}{2} \tr ( M^\dagger M)
\end{equation}
which is manifestly invariant under~\eqref{CStransSM} so that
$ V_{SM} $ will be CT invariant.
It is easy to see, however, that the CS is explicitly
violated in the SM by both the hypercharge gauge interactions
and the Yukawa couplings; so the CS is but an approximate symmetry in the SM.
Note that when the CS-violating coefficients are set to zero
all massive vector bosons are mass degenerate 
(corresponding to $ \rho = 1 $) to all orders in perturbation theory
\cite{Sikivie:1980hm, Peskin:1995ev}.

It is easy to see that, even at tree-level, $\rho=1$ cannot be realized
naturally in an extended scalar sector unless the
scalar multiplets belong to a specific set of
isospin representations. 
The singlets and isodoublets
are the simplest of these ``$\rho$-safe'' representations;
hereafter we will focus on  isodoublet extensions of the SM.
It is well known~\cite{Sikivie:1980hm, Toussaint:1978zm, Pomarol:1993mu}  
that even for two scalar isodoublets,
in general, there exist potentially large radiative corrections to $\rho-1$ 
proportional to the squares of the scalar masses. A remedy has 
also been proposed~\cite{Sikivie:1980hm, Toussaint:1978zm, Pomarol:1993mu} 
through two generalizations of the CT 
to the two-Higgs-doublet model (THDM)~\footnote{
Recently the subject has been resurrected in~\cite{Gerard:2007kn}.}. 

In this note we will revisit this issue and 
derive basis-independent conditions for the
CS for the THDM potential using both the conventional 
approach and 
the bilinear formalism of~\cite{Nagel:2004sw, Maniatis:2006fs}, 
which allows to illustrate the CT in a transparent way. 
We also discuss the relation between CT and CP invariance.

\section{The custodial symmetry}

The most general potential for the THDM may be written in terms of the 
following doublets carrying the same hypercharge:
\begin{equation}
\phi_1 = 
\begin{pmatrix}
\phi_1^+ \\ \phi_1^0
\end{pmatrix},
\qquad
\phi_2 = 
\begin{pmatrix}
\phi_2^+ \\ \phi_2^0
\end{pmatrix},
\end{equation}
Then the potential reads~\cite{Haber:1993an}
\begin{equation}
\begin{split}
V =& \phantom{+}
m_{11}^2 (\phi_1^\dagger \phi_1) +
m_{22}^2 (\phi_2^\dagger \phi_2) 
\\
&
- m_{12}^2 (\phi_1^\dagger \phi_2) -
(m_{12}^2)^* (\phi_2^\dagger \phi_1)
\\
&
+\frac{1}{2} \lambda_1 (\phi_1^\dagger \phi_1)^2 
+ \frac{1}{2} \lambda_2 (\phi_2^\dagger \phi_2)^2 
+ \lambda_3 (\phi_1^\dagger \phi_1)(\phi_2^\dagger \phi_2)
\\ 
&
+ \lambda_4 (\phi_1^\dagger \phi_2)(\phi_2^\dagger \phi_1)
+ \frac{1}{2} [\lambda_5 (\phi_1^\dagger \phi_2)^2 + \lambda_5^* 
(\phi_2^\dagger \phi_1)^2]
\\ 
&
+ [\lambda_6 (\phi_1^\dagger \phi_2) + \lambda_6^* 
(\phi_2^\dagger \phi_1)] (\phi_1^\dagger \phi_1) \\
&
+ [\lambda_7 (\phi_1^\dagger 
\phi_2) + \lambda_7^* (\phi_2^\dagger \phi_1)] (\phi_2^\dagger \phi_2)\;,
\label{V_fields}
\end{split}
\end{equation}
with $m_{11}^2$, $m_{22}^2$, $\lambda_{1,2,3,4}$ real and 
$m_{12}^2$, $\lambda_{5,6,7}$ complex.

For studies of the CS within the THDM 
it is convenient to introduce the following set of 
matrixes
\begin{equation}
M_{ij} \equiv (\tp_i,\phi_j)=
\begin{pmatrix}
\phi^{0\star}_i & \phi^+_j \\ -\phi^-_i & \phi^0_j
\end{pmatrix} 
\end{equation}
where $i=1,2$ refers to the scalar doublets. It is easy to see that all bilinears $\phi_i^\dagger \phi_j$ 
can be expressed in terms of $M_{11}$ and $M_{22}$, or in terms of $M_{12}$. Therefore
the scalar potential (\ref{V_fields}) could also be written using $M_{11}$ and $M_{22}$ 
or  $M_{12}$.

The following two versions of the CT for THDM have been considered in the literature~\cite{Pomarol:1993mu}:
\bit
\item Type I: 
In this case it is useful to express the potential in terms of  $M_{11}$ and $M_{22}$. 
The transformation is a straightforward generalization of (\ref{CStransSM}):
\beq
M_{ii} \stackrel{CT_{I}}{\lra} M_{ii}^\prime = L M_{ii} R^\dagger \for  i=1,2
\label{cust-symm-I}
\eeq
\item Type II: 
For this version of the CT, considered in~\cite{Pomarol:1993mu},
it is convenient to express the potential using $M_{12}$ only.
The corresponding CT reads:
\begin{equation}
M_{21} \stackrel{CT_{II}}{\lra} M_{21}^\prime = L M_{21} R^\dagger \;,
\label{cust-symm-II}
\end{equation}
\eit
where $L$ and $R$ belongs to $SU(2)_L$ and $SU(2)_R$, respectively. Note that one cannot simultaneously 
have invariance under \eqref{cust-symm-I} and  \eqref{cust-symm-II} since, for example, 
the second case mixes $ \phi_1 $ and $ \phi_2 $ while the first one does not.

It is worth mentioning here that, since both Higgs-boson doublets carry the same quantum numbers,
physical content of the model we are considering cannot depend on a choice of 
the basis adopted for the scalar doublet fields $(\phi_1,\phi_2)$.
Nevertheless, the form of the Lagrangian obviously changes by 
a change of basis; also the form of custodial transformation will change if we change the basis.
In what follows we will investigate 
consequences of such a unitary basis transformation:
\beq
\phi_i \to \hp_i = \sum_j U_{ij}\phi_j \for i=1,2
\label{phi-basis-trans}
\eeq
where $U \in U(2)$. This rotation implies the following change for
$M_{ij}$
\begin{multline}
M_{ij}  \equiv (\tp_i,\phi_j)  \to  \hat M_{ij} \equiv (\hat\tp_i,\hat\phi_j) = \\
\left( \sum_k U_{ik}^\star \tp_k, \; \sum_l U_{jl} \phi_l \right) = \\
\frac12 \sum_{kl} M_{kl}
\left(
\baa{cc} U_{ik}^\star & 0 \\ 0 & U_{jl} \eaa
\right) .
\label{M-basis-trans}
\end{multline}
Note that above the sum stands in front of a product of matrices, so that
the elements of summed matrices are correlated.
From~\eqref{M-basis-trans} we can determine the form of the
CT in the new basis; for example, for type I~\eqref{cust-symm-I} we obtain,
\beq
\hat M_{ij} \stackrel{CT_{I}}{\lra}  \frac12 
\sum_{k} 
\sum_{op}
L \hat M_{op}
\left[
\left(
\baa{cc} U_{ok} & 0 \\ 0 & U_{pk}^\star \eaa
\right)
R^\dagger
\left(
\baa{cc} U_{ik}^\star & 0 \\ 0 & U_{jk} \eaa
\right)
\right] .
\label{CS-gen-new-bas} 
\eeq
%

\section{CT in terms of gauge invariant bilinear}

Due to gauge invariance the doublets in the potential~\eqref{V_fields} can
only appear in bilinear form, that is, in terms of~$(\phi_i^\dagger \phi_j)$.
It turns out that it is very convenient to discuss
CS using the bilinears instead of the fields themselves. There are just four independent bilinears
that can be combined as follows~\cite{Nagel:2004sw,Maniatis:2007vn}
\begin{equation}
K_0 = \phi_1^\dagger \phi_1 + \phi_2^\dagger \phi_2,
\qquad 
\tvec{K} =
\left(
\begin{array}{c}
\phi_1^\dagger \phi_2 + \phi_2^\dagger \phi_1 \\
i \phi_2^\dagger \phi_1 -  i \phi_1^\dagger \phi_2 \\
\phi_1^\dagger \phi_1 - \phi_2^\dagger \phi_2
\end{array}
\right).
\label{connect_4}
\end{equation}
Any choice of $K_0 \ge 0$ and $\tvec{K}^2 \le K_0^2$ fix
the doublets modulo a gauge transformation. 
The potential (\ref{V_fields}) can be expressed in terms of $K_0$ and $\tvec{K}$
in a very compact and suggestive way
\begin{equation}
\label{VK}
V = \xi_0 K_0 + \tvec{\xi}^\trans \tvec{K} + \eta_{00} K_0^2 + 2 K_0 \tvec{\eta}^\trans \tvec{K} + \tvec{K}^\trans E \tvec{K}
\end{equation}
with the following real parameters
\begin{equation}
\begin{split}
\label{Kpara}
&\xi_0, \quad \eta_{00}, \quad
\tvec{\xi}= \begin{pmatrix} \xi_1\\ \xi_2\\ \xi_3 \end{pmatrix}, \quad
\tvec{\eta}= \begin{pmatrix} \eta_1\\ \eta_2\\ \eta_3 \end{pmatrix}, \\
&E=E^\trans=
\begin{pmatrix}
E_{11} & E_{12} & E_{13}\\
E_{12} & E_{22} & E_{23}\\
E_{13} & E_{23} & E_{33}
\end{pmatrix}
\end{split}
\end{equation}
which have the following expressions in terms of the original parameters in \eqref{V_fields}:
\begin{equation}
\begin{split}
\xi_0&=\frac{1}{2}
(m_{11}^2+m_{22}^2)\;, 
\quad
\tvec{\xi}=\frac{1}{2}
\left(
\begin{array}{c}
- 2 \textrm{Re}(m_{12}^2) \\
2 \textrm{Im}(m_{12}^2)\\
 m_{11}^2-m_{22}^2
\end{array}
\right),\\
\eta_{00}& =
\frac{1}{8}(\lambda_1 + \lambda_2) + \frac{1}{4}\lambda_3\;,
\quad
\tvec{\eta}=\frac{1}{4}
\left(
\begin{array}{c}
\textrm{Re}(\lambda_6+\lambda_7)\\
-\textrm{Im}(\lambda_6+\lambda_7)\\ 
\frac{1}{2}(\lambda_1 - \lambda_2)
\end{array}
\right),
\label{connect_2}
\end{split}
\end{equation}
\begin{equation*}
E = \frac{1}{4}
\left(
\begin{array}{ccc}
\lambda_4 + \textrm{Re}(\lambda_5) & 
-\textrm{Im}(\lambda_5) & 
\textrm{Re}(\lambda_6-\lambda_7) \\
-\textrm{Im}(\lambda_5) & 
\lambda_4 - \textrm{Re}(\lambda_5) & 
-\textrm{Im}(\lambda_6-\lambda_7) \\ 
\textrm{Re}(\lambda_6-\lambda_7) & 
-\textrm{Im}(\lambda_6 -\lambda_7) & 
\frac{1}{2}(\lambda_1 + \lambda_2) - \lambda_3
\end{array}
\right).
\end{equation*}
It is easy to see that the transformations of $K_0$ and $\tvec{K}$ under
a change of basis~\eqref{phi-basis-trans} read 
\begin{equation}
\label{eq12}
\begin{split}
&K_0 \to K'_0 = K_0,\\
&\tvec{K} \to \tvec{K}' = R(U)  \tvec{K}\;,
\end{split}
\end{equation}
where the matrix $R(U) \in SO(3)$ is given in terms of $U$ through
\begin{equation}
\label{eqUR}
U^\dagger \sigma^a U = R_{ab}(U)\,\sigma^b.
\end{equation}
The effect of a change of basis~\eqref{eq12}
corresponds to the following change of the potential parameters
\begin{equation}
\begin{split}
\label{eqparabasis}
\tvec{\xi} &\to \tvec{\xi}' = R(U)^\trans \tvec{\xi} \;, \\
\tvec{\eta} &\to \tvec{\eta}' = R(U)^\trans \tvec{\eta} \;, \\
 E &\to E' = R(U)^\trans E R (U)\;.
\end{split}
\end{equation}
Now we turn to the description of the CS in terms of these parameters.

%
\subsection*{Custodial transformation of type I}

For the CT of type~I~\eqref{cust-symm-I} it is convenient to express
$ K_0, ~\tvec{K} $ in terms of $M_{ii}$ with $i=1,2$:
\begin{equation}
\begin{split}
& K_0= \frac12 \tr (M_{11}^\dagger M_{11}+M_{22}^\dagger M_{22}) ,\\
& K_1= \tr (M_{11}^\dagger M_{22}) ,\\
& K_2= (-i) \tr (M_{11} \tau_3 M_{22}^\dagger) ,\\
& K_3= \frac12 \tr (M_{11}^\dagger M_{11}-M_{22}^\dagger M_{22}) .
\label{K-def-I}
\end{split}
\end{equation}
whence the type I CT corresponds to
\begin{equation}
\begin{split}
\label{K-symm-I}
\text{CT}_I:\quad &K_{0,1,3} \to K_{0,1,3} ,\\
\qquad &K_2 \to (-i) \tr\left[ M_{11} (R^\dagger \tau_3 R) M_{22}^\dagger\right].
\end{split}
\end{equation}
The invariance of the potential (\ref{VK}) under (\ref{K-symm-I}) restricts
its parameters as follows
\begin{equation}
\label{paraI}
\tvec{\eta}=
\begin{pmatrix} \cdot \\ 0 \\ \cdot \end{pmatrix}, \quad
\tvec{\xi}=
\begin{pmatrix} \cdot \\ 0 \\ \cdot \end{pmatrix}, \quad
E=
\begin{pmatrix} 
\cdot & 0 & \cdot \\
0 &  0  & 0 \\
\cdot & 0 & \cdot
\end{pmatrix} ,
\end{equation}
where the dots denote arbitrary entries. 
{\em The THDM will be symmetric under this CT if
and only if there exists a basis rotation after which
the potential parameters take the form \eqref{paraI}.}

Using \eqref{connect_2} we can express the potential~\eqref{VK} 
with
parameters~\eqref{paraI} in terms of the original doublets:
\begin{equation}
\label{paraIconv}
\begin{split}
V &=
m_{11}^2 (\varphi_1^\dagger \varphi_1) +
m_{22}^2 (\varphi_2^\dagger \varphi_2) \\
&
- \re(m_{12}^2) [(\varphi_1^\dagger \varphi_2) + (\varphi_2^\dagger \varphi_1)]
\\
&
+\frac{1}{4} (\lambda_1+\lambda_2) 
[(\varphi_1^\dagger \varphi_1)^2 + (\varphi_2^\dagger \varphi_2)^2]
+ \lambda_3 (\varphi_1^\dagger \varphi_1)(\varphi_2^\dagger \varphi_2)
\\ 
&
+ \frac{1}{2}(\lambda_4+\re(\lambda_5)) 
[(\varphi_1^\dagger \varphi_2)+(\varphi_2^\dagger \varphi_1)]^2
\\ 
&
+ \re(\lambda_6) [(\varphi_1^\dagger \varphi_2) + (\varphi_2^\dagger \varphi_1)] 
(\varphi_1^\dagger \varphi_1) 
\\
&
+ \re(\lambda_7) [(\varphi_1^\dagger \varphi_2) + (\varphi_2^\dagger \varphi_1)] 
(\varphi_2^\dagger \varphi_2)\;.
\end{split}
\end{equation}
This potential is invariant under the CT of type I and
matches the expression in~\cite{Pomarol:1993mu}.


\subsection*{Custodial transformation of type II}

In this case it is 
useful to express $K_0$ and $\tvec{K}$ in terms of $M_{21}$ only:
\begin{equation}
\begin{split}
& K_0= \tr (M_{21}^\dagger M_{21}) ,\\
& K_1= 2 \re (\det M_{21}^\dagger) ,\\
& K_2= -2 \im (\det M_{21}) ,\\
& K_3= -\tr (M_{21}\tau_3 M_{21}^\dagger) 
\label{K-def-II}
\end{split}
\end{equation}
which then transform as follows:
\begin{equation}
\begin{split}
\text{CT}_{II}:\quad &K_{0,1,2} \to K_{0,1,2} , \\
&K_3 \to -\tr\left[ M_{21} (R^\dagger \tau_3 R) M_{21}^\dagger\right] \;.
\label{K-symm-II}
\end{split}
\end{equation}
It follows that in order for the potential to be invariant under this CT
the parameters take the form
\begin{equation}
\label{paraII}
\tvec{\eta}=
\begin{pmatrix} \cdot \\ \cdot \\ 0 \end{pmatrix}, \quad
\tvec{\xi}=
\begin{pmatrix} \cdot \\ \cdot \\ 0 \end{pmatrix}, \quad
E=
\begin{pmatrix} 
\cdot & \cdot & 0 \\
\cdot & \cdot & 0 \\
0 & 0 & 0
\end{pmatrix} .
\end{equation}
Again we can use \eqref{connect_2} to
write the potential in terms of the two doublets as 
presented in~\cite{Pomarol:1993mu}:
\begin{equation}
\label{paraIIconv}
\begin{split}
V &=
\frac{1}{2}(m_{11}^2+m_{22}^2) 
[(\varphi_1^\dagger \varphi_1) + (\varphi_2^\dagger \varphi_2)] \\
&
-m_{12}^2 (\varphi_1^\dagger \varphi_2) -
(m_{12}^2)^* (\varphi_2^\dagger \varphi_1)
\\
&
+\frac{1}{4} (\lambda_1+\lambda_2+2\lambda_3)
[ (\varphi_1^\dagger \varphi_1)+ (\varphi_2^\dagger \varphi_2) ]^2
\\ 
&
+ \lambda_4 (\varphi_1^\dagger \varphi_2)(\varphi_2^\dagger \varphi_1)
+ \frac{1}{2} [\lambda_5 (\varphi_1^\dagger \varphi_2)^2 
+ \lambda_5^* (\varphi_2^\dagger \varphi_1)^2]
\\ 
&
+ (\im(\lambda_6)+\im(\lambda_7))i 
[(\varphi_1^\dagger \varphi_2)-(\varphi_2^\dagger \varphi_1)](\varphi_1^\dagger \varphi_1)\\
&
+ (\im(\lambda_6)+\im(\lambda_7))i 
[(\varphi_1^\dagger \varphi_2)-(\varphi_2^\dagger \varphi_1)](\varphi_2^\dagger \varphi_2)\;.
\end{split}
\end{equation}
Note that the parameters are in general complex in this case.

Here we immediately see the advantage of the bilinear formalism:
while the potentials in conventional notation, \eqref{paraIconv} and
\eqref{paraIIconv} look quite different, in terms of bilinears
$ K_0~ \tvec{K}$ they are very similar
(compare \eqref{paraI} with \eqref{paraII}).
In the next section we will show that both
potentials are related by a simple basis transformation.

\subsection{Equivalence of the two types of custodial transformation}

Here we will show that  type I \eqref{cust-symm-I} and  type II \eqref{cust-symm-II}
CT are equivalent, that is, these are the same transformation expressed in different bases.

In the bilinear formalism this is evident:
the parameters ~\eqref{paraI}
and~\eqref{paraII}, corresponding to type I and II CT
are related by a change of basis~\eqref{eqparabasis} with
\begin{equation}
\begin{split}
R_{I \to II}^{(1)} &= 
\begin{pmatrix}
\sin\alpha & 0 & -\cos\alpha \\
\cos\alpha & 0 &  \sin\alpha \\
0 & -1 & 0
\end{pmatrix} 
\lsp {\rm or} \\
R_{I \to II}^{(2)} &= 
\begin{pmatrix}
\sin\alpha & 0 &  \cos\alpha \\
\cos\alpha & 0 & -\sin\alpha \\
0 & 1 & 0
\end{pmatrix} .
\label{R12}
\end{split}
\end{equation}
Then solving the equation~\eqref{eqUR} one finds that
the corresponding $U$ is 
\begin{equation}
\label{U-res}
U_{I \to II} = \frac{e^{i\gamma}}{\sqrt{2}}  
\begin{pmatrix}
e^{-i\varphi} & \pm i e^{-i\varphi}  \\  
e^{+i\varphi} & \mp i e^{+i\varphi} 
\end{pmatrix} 
\end{equation}
where $\gamma$ is an undetermined phase 
(the method involving bilinears is not sensitive to an overall phase, which
can always be absorbed by a hypercharge transformation of the doublets) 
and upper and lower signs correspond to
$R_{I \to II}^{(1)}$ with $\varphi=-\alpha/2 + \pi/2$, and $R_{I \to II}^{(2)}$
with $\varphi=-\alpha/2 $, respectively. 
The rotation $U_{I\to II}$ can, of course, also be obtained from
(\ref{CS-gen-new-bas}) by requiring that the $ \hat\phi_i $
transform according to (\ref{cust-symm-II}) whenever the $ \phi_i $
transform according to (\ref{cust-symm-I}).

We close this section with a comment on the 
equivalence of these two types of CT in a realistic 
theory that also contains fermions. For the clarity of the argument, 
let us first assume that Yukawa interactions are absent.
Then consider the following two versions (I and II) of THDM: I) A model with 
the potential $V_I(\vec \phi)$ and II) a model with the potential 
$V_{II}(\vec \phi) := V_I(U \vec \phi)$ (equivalent 
to $V_I(\vec\phi)$, just written in a different basis);
the potentials (\ref{paraIconv}) and (\ref{paraIIconv}) serve as an illustration of these two versions,
as they are related by the unitary transformation (\ref{U-res}).
As long as Yukawa interactions are not present the two models are physically identical, they
differ only by a field redefinition.
Now let us switch on Yukawa interactions of the same form in both 
Lagrangians, $\vec \phi \bar \psi \vec\Gamma_Y \psi$, 
so that $\lcal_I(\vec\phi) = \cdots - V_I(\vec \phi) + \vec\phi \bar \psi \vec\Gamma_Y \psi + \cdots$ while 
$\lcal_{II}(\vec\phi) = \cdots - V_I(U \vec \phi) + \vec\phi \bar \psi \vec\Gamma_Y \psi + \cdots$. Obviously,
now $\lcal_I$ and $\lcal_{II}$
are no longer equivalent; they differ by the potentials. 
However it is interesting to realize that where the difference between them is 
located is a matter of basis choice: the basis change $\vec \phi \to U^{-1}\vec \phi $ performed upon $\lcal_{II}$ would shift the
difference from the potentials to the Yukawa interactions. It is then interesting 
to note that in the perturbative expansion only those processes
are sensitive to the difference between the two versions that incorporate 
both Yukawa couplings {\it and} couplings that emerge from scalar 
potentials (so e.g. scalar masses). For instance, 
the vector-boson vacuum polarizations would be exactly the same in both models
at 1-loop (but not in higher orders).

The equivalence (by a basis transformation) between the two types of 
the transformations that we have found above clearly shows a need for a basis independent 
formulation of an invariance under the CT. That issue is discussed in the
next section.

\subsection{Basis independent conditions for CS}

The two types of CT considered above are related by a change of basis
and are therefore equivalent, but it would clearly be
desirable to 
have a basis-independent set of conditions 
which insure that a scalar potential is invariant under CT.
In terms of the bilinear coefficients 
in (\ref{VK}) these conditions are the following 
\bea
\label{custcond}
 E.\vv &=& 0  \cr
\tvec\xi.\vv = \tvec\eta.\vv &=&0 
\eea
for some $\vv \neq 0$.
In order to prove the assertion we note that these conditions
are basis independent since the first one is equivalent to requiring
det$(E)=0 $.
This means that if the conditions~\eqref{custcond} are
satisfied in one basis, they are satisfied in any basis.
Therefore it is sufficient to show that~\eqref{custcond}
are necessary and sufficient conditions for a custodial
symmetry in a specific basis, for instance the one defined by the parameters
\eqref{paraII}.

First we have to show that~\eqref{paraII} imply~\eqref{custcond},
which is immediate: $ \vv = (0,0,1) $ is the zero eigenvector of $E$
in \eqref{paraII}, and \vv\ is indeed orthogonal to $ \tvec\xi,\tvec\eta$.
Now, assume~\eqref{custcond}, then, since $E$ is symmetric and has
one zero eigenvalue, we can choose a basis where
$E = \hbox{diag}(E_1,E_2,0) $, so that we can take $ \vv = (0,0,1)$,
and this will be orthogonal to $ \tvec\xi,\tvec\eta$ only if
both these vectors take the form $(\cdot,\cdot,0) $; it follows
that there is a basis where \eqref{custcond} imply~\eqref{paraII}.

\section{CT versus CP symmetry}

In the bilinear formalism it is easily seen that
custodial symmetry and CP symmetry are closely related.
First we recall the CP transformation of the
doublets, which is defined by
\begin{equation}
\label{CP}
\text{CP}:\quad \varphi_i(x) \rightarrow \varphi_i^*(x'), \qquad i=1,2.
\end{equation}
Here we have explicitly written the argument of the fields,
since the argument is changed under the CP transformation, that is,
we have $x' = (x^0, -\tvec{x})$.
Applying~\eqref{CP} to the bilinears~\eqref{connect_4} we
see that a
CP transformation is a reflection on the 1--3 plane -- in addition
to the parity transformation for the field argument~\cite{Maniatis:2007vn}:
\begin{equation}
\begin{split}
\label{KCP}
\text{CP}:\quad &K_{0,1,3}(x) \to K_{0,1,3}(x') ,\\
&K_2(x) \to -K_2(x')
\end{split}
\end{equation}
We recognize that like in the type I of the CT~\eqref{K-symm-I}
only the bilinear $K_2$ transforms nontrivially under CP. 
We can now easily give 
the Higgs potential, invariant under~\eqref{KCP}, which
has to have the following parameters~\cite{Maniatis:2007vn}
\begin{equation}
\label{paraCP}
\tvec{\xi}=
\begin{pmatrix} \cdot \\ 0 \\ \cdot \end{pmatrix}, \quad
\tvec{\eta}=
\begin{pmatrix} \cdot \\ 0 \\ \cdot \end{pmatrix}, \quad
E=
\begin{pmatrix} 
\cdot & 0 & \cdot \\
0 &  \cdot  & 0 \\
\cdot & 0 & \cdot
\end{pmatrix}.
\end{equation}
By a comparison with~\eqref{paraI} we find
that the only difference is the central entry of the matrix $E$.
We thus can state that any Higgs potential, invariant under custodial
symmetry, is automatically invariant under the CP transformation.
Note that the opposite is in general not true.
This result holds in any basis, since we can for any custodial
symmetric model -- by a change of basis -- go to the parameterization~\eqref{paraI}.
We can also find this result by a comparison of the basis-independent
conditions given for the CP transformation in~\cite{Maniatis:2007vn}
and for the custodial symmetry given in~\eqref{custcond}.

It is worth noticing that the requirement of breaking $SU(2)_L\times SU(2)_R$
down to $\su2$ implies $v_i=v_i^\star$, in other words the possibility
of spontaneous CP violation is also eliminated by the requirement of 
invariance under the CT.

\section{Comments and conclusions}

The custodial symmetry in the SM is respected by the Higgs potential
implying no corrections to the $\rho$ parameter which grow as $\propto m_h^2$. 
However in the Two-Higgs-Doublet Model the potential in
general does not respect this symmetry. In this paper, employing the
bilinear formalism,  we have formulated basis independent
conditions~\eqref{custcond} which allow for an easy verification of 
the custodial symmetry of a scalar potential. 
We have also shown that, {\it as long as Yukawa interactions are 
irrelevant}, two types of custodial symmetry
for THDM discussed in the literature are equivalent; they just 
differ by a choice of basis. We have also clarified relations between
the custodial symmetry and CP; it has been shown that any potential which
is symmetric under custodial symmetry is also invariant under CP.

\vspace{.5in}
\acknowledgments

M.M.~wants to thank O.~Nachtmann for
very helpful discussions.
This work is supported in part by the Ministry of Science and Higher
Education (Poland) as research project N~N202~006334 (2008-11). 


\end{document}